\documentstyle[12pt,here,epsf,rotate]{article}
\newcommand{\bel}[1]{\begin{equation}\label{#1}}
\newcommand{\ee}{\end{equation}}

\def\IM{{\rm Im}}
\def\om{\omega}
\def\omd{\varpi_D}
\newcommand{\pol}[1]{\omega^{(#1)}}
\def\oscsigmaqq{\Sigma^{\rm osc}_{qq}}

\def\oscsigmapp{\Sigma^{\rm osc}_{pp}}
\def\dissip{\chi^{\prime\prime}}
\def\Om{\Omega}
\def\eps{\varepsilon}
\def\eqsigmaqq{\Sigma^{\rm eq}_{qq}}

\def\chidru{\chi^{\rm dru}}
\def\xvi{{\hat x}_i}
\def\pvi{{\hat p}_i}
\newcommand{\rfc}[1]{\chi_{{\rm coll}} (#1)}
\newcommand{\ham}[1]{\hat H(\xvi,\pvi,#1)}
\newcommand{\ffield}[1]{ \hat F(\xvi,\pvi,#1)}
 
\def\fmb{ \langle \hat F \rangle }
\newcommand{\fm}[1]{ \langle \hat F \rangle _{#1}}

\author{H.~Hofmann 
\thanks{e-mail: hhofmann@.physik.tu-muenchen.de}
\thanks{http://www.physik.tu-muenchen.de/tumphy/e/T36/hofmann.html}\\
        Physik-Department, TU M\"unchen, D-85747 Garching\\
        and\\
        D.~Kiderlen\\
        National Superconducting Cyclotron Laboratory\\
        Michigan State University, East Lansing, Michigan 48824
        \vspace{0.8cm}}
\title{A self-consistent treatment of damped motion for stable and
unstable collective modes\thanks{Dedicated to Professor Richard Lemmer
on the occasion of his 65th birthday}
  \vspace{0.5cm}}\date{\today}
\begin{document}

\maketitle

\begin{abstract}
\noindent
We address the dynamics of damped collective modes in terms of first
and second moments. The modes are introduced in a self-consistent
fashion with the help of a suitable application of linear response
theory.  Quantum effects in the fluctuations are governed by diffusion
coefficients $D_{\mu\nu}$. The latter are obtained through a
fluctuation dissipation theorem generalized to allow for a treatment of
unstable modes. Numerical evaluations of the $D_{\mu\nu}$ are
presented. We discuss briefly how this picture may be used to describe
global motion within a locally harmonic approximation.
Relations to other methods are discussed, like "dissipative tunneling",
RPA at finite temperature and generalizations of the "Static Path
Approximation".

\end{abstract}

\noindent
 \vspace*{1cm}
\centerline{PACS: 5.40+j, 5.60+w, 24.10 Pa, 24.60-k}
 \vspace*{1cm}
\section{Introduction}

The interplay of dissipation and quantum effects has been a long
standing challenge to transport theory, in particular the question of
the influence of damping on quantum tunneling. Convincing answers have
only be found in the 80'ies by clarifying demanding problems of
principal nature with the help of functional integrals, see e.g.
\cite{calleg} 
- \cite{graschring}. 
In these approaches one first derives effective
actions for global motion from which one may then deduce expressions
for the decay rate of a meta-stable state. In contrast to this
procedure, we like to look at the problem from the point of view of
transport equations such as the one of Kramers \cite{kram} but extended
to include quantum effects. By its very nature, such a transport
equation reflects local properties of collective motion as displayed
through the propagation in collective phase space. In the following we
like to concentrate on situations where damping cannot simply be
treated perturbatively. For the case of stable modes, suggestions of
possible quantal equations have been made for instance in \cite{hosaoc}
and \cite{dekquatra}. The possibility of generalizing to unstable modes
has first been described in
\cite{hofing}.

Different to those papers mentioned previously which deal with
dissipative tunneling and which exploit functional integrals, we like
to base our analysis on systems where collective variables  are
introduced {\it self-consistently}. For Hamiltonian dynamics, this
implies that in the Hamiltonian
\bel{tothamdef}
{\cal H} = H_{\rm intr} + H_{\rm coupl} + H_{\rm coll} 
\ee
for the total system both the "collective" part $H_{\rm coll}$ as well
as the coupling term are bound to be  functionals of the dynamical
variables of the intrinsic system. One prime example is the electron
gas for which D. Bohm and D. Pines have been able to deduce such a
Hamiltonian in \cite{bopi}. Another example is nuclear collective
motion for which an adapted version of the Bohm-Pines method can be
found in \cite{hoso}, see also \cite{hofrep}. Although later on we are
going to concentrate on systems for which such a Hamiltonian
(\ref{tothamdef}) exists it may be mentioned that our method of
accounting for self-consistency in the fluctuating force can be
generalized to other cases, the most prominent example being a
situation in which the dynamics is described by the Landau-Vlasov
equation with (collision term) (see \cite{kidhofpl}).  For stable modes
this method may in some sense be considered a generalization of the
suggestion made by Landau \cite{lalifhyd} (see
\cite{kidself}). 

Evidently problems like the motion from a meta-stable minimum across a
potential barrier are manifestly non-linear in at least one of the
collective degrees of freedom. Moreover, there are systems for which
the intrinsic sub-system and its coupling to the collective part change
along the dynamical path, thus prohibiting to linearize this coupling
in the bath variables. In such a situation, as it is given for example
in nuclear fission, the technique of functional integration is not
useful anymore as the internal degrees of freedom can no longer be
integrated out. It has been argued that it might be possible to handle
such systems by a {\it locally harmonic approximation (LHA)} (see e.g.
\cite{berlinhi},
\cite{hofnorps}, \cite{hofrep}). In this method it is not necessary to
restrict the dynamics of the internal degrees of freedom to a system of
(perturbed) oscillators.  Certainly, as an essential intermediate step,
there will appear the dynamics in linearized version for the {\it
collective degrees of freedom}. As already demonstrated by Kramers, in
lucky cases it may even suffice to have information on the dynamics at
the potential minimum and at the barrier, the latter mode being
unstable.  In this paper we shall address specifically these problems
and study the influence of dissipation on average motion as well as on
the dynamics of the second moments (fluctuations).  This will be
achieved within the framework of linear response theory. The latter
does not only allow one to treat average motion (in a linearized
fashion), it opens an ideal playground for deducing the diffusion
coefficients which govern the dynamics of the second moments. To be
able to treat instabilities as well, it will turn out necessary to
perform suitable analytical continuations.  Manipulations of the kind
which will be needed here have been used before in \cite{ingold},
\cite{ankgraingpa}, but, as mentioned, not in connection with a
transport equation and, hence, diffusion coefficients.  As we shall see
the numerical computations of these coefficients will very naturally
hint to limits of the LHA at smaller temperatures.

\section{Brief review of average motion}

In this paper we would like ourselves to restrict to a situation in
which the dynamics of interest can be described by the one quantity
$\hat F$. For average motion this means looking at the time evolution
of the expectation value $\fmb_t$. To start with a simple situation let
us suppose we are given a Hamiltonian of the type
\bel{twobodham}
\hat H^{(2)}= \hat H_0  + {k\over2} \hat F \hat F
\ee
with some coupling constant $k$, without bothering for the moment how
it has come about. Later on we are going to describe how this form may
derived from more general considerations and how the $k$ may be deduced
from static properties of the system, in which case it will depend on
the actual state of the intrinsic system. Let us suppose further that
we are interested in harmonic vibrations. As may be well known, their
frequencies and strengths can be deduced from the following response
function
\bel{chicoll}
\rfc{\om} = {\chi(\om)\over 1+k\chi(\om)}
\ee
obeying the following definition
\bel{qext}
\delta \fmb_\om   = - \rfc{\om} f_{\rm ext}(\om) 
\ee
The $\delta \fmb_\om$ measures the Fourier transform of the deviation
of $\fmb_t$ from the equilibrium value $\fmb_0$, as it follows from the
response of the system in linear order to an external field $ f_{\rm
ext}(\om)$, with the coupling being given by $\delta \hat H= f_{\rm
ext} \hat F$. The $\chi(\om)$ is the Fourier transform of the response
function defined by
\bel{defresmic}
\widetilde{\chi} (t-s)= \Theta(t-s) {i\over \hbar} 
             {\rm tr}\,\left(\hat{\rho}^0_{\rm qs}
             \Big[\hat F^{I}(t),\hat F^{I}(s)\Big] \right) 
\ee
with the $\hat{\rho}_{\rm qs}^0$ representing the equilibrium density
operator associated to the Hamiltonian $\hat H_0$. Notice please that
this function is governed entirely by {\it intrinsic} properties,
different to the $\rfc{\om}$ which contains information on collective
motion.

We should expect average dynamics to be closely related to the mean
field approximation. Indeed, for the latter one readily verifies the
following Hamiltonian
\bel{hammeanf}
\hat H^{(2)}_{\rm mf}= \hat H_0 + (Q-Q_0) \hat F
\ee
where the abbreviation
\bel{selfcons}
k \fm{t}= Q-Q_{0} 
\ee
has been used. One may now simply apply the Clausius-Mosotti
construction to derive the form  (\ref{chicoll}) from (\ref{qext}) and
(\ref{selfcons}).

For the following it will show convenient to add to the Hamiltonian
$\hat H^{(2)}_{\rm mf}$ of (\ref{hammeanf}) a c-number term and thus to
introduce a  $\ham{Q}$ like
\bel{hamexp} 
\ham{Q}=\ham{Q_0} + (Q-Q_0)\ffield{Q_0} + {1\over 2}\left(Q-Q_0\right)^2 
\big\langle{\partial ^2 \ham{Q}\over \partial Q^2} \big\rangle_{Q_0} 
\, .
\ee
Identifying $\ham{Q_0}=\hat H_0$ and the expectation value on the very
right with  
\bel{defcoupcon}
 -k^{-1} = 
\big\langle{\partial ^2 \ham{Q}\over \partial Q^2} \big\rangle_{Q_0} 
\ee
one sees that the total energy of the system  $E_{\rm tot} = 
\big\langle\hat H^{(2)}_{\rm mf}\big\rangle
 -  {k\over2} \fmb \fmb  $ can simply be expressed as the expectation
value $\big\langle\ham{Q}\big\rangle$ of the one-body Hamiltonian
$\ham{Q}$. 

This discussion has demonstrated the intimate relation between a
Hamiltonian of the form  (\ref{hamexp}) and the one of
(\ref{twobodham}) involving a schematic two-body interaction. On the
level of the mean field approximation the latter becomes equivalent to
the former. However, the $\hat H^{(2)}$ of (\ref{twobodham}) allows one
to treat fluctuations in $\hat F$, or what will turn out more
convenient in $Q-Q_0$, about its average value. Indeed, later on we
will be exploiting exactly this feature when we are going to quantize
the "collective variable" $Q$. Conversely, as often mean field
approximations are more easily accessible numerically than solutions of
the full Schr\"odinger or Heisenberg equations (of the many body
problem), one may start from such a $\ham{Q}$ and construct the
effective two body interaction appearing in $\hat H^{(2)}$.

A prominent example from nuclear physics is given when one tries to
simulate the mean field approximation via the deformed shell model. In
that case the $Q$ is introduced as a c-number variable specifying the
shape of the nucleus. The latter then appears both in the single
particle potential as well as in the liquid drop energy. To account for
the latter is necessary because otherwise one would not be able to
calculate reliably the total energy of the system. This can be done
with the help of the Strutinsky renormalization. How one may then
proceed to construct a Hamiltonian of the type $\ham{Q}$ has been
worked out in detail in \cite{sije}.

It is more than conceivable that one might start from even more
microscopic theories like the one of Hartree-Fock involving effective
forces or the Relativistic Mean Field theory. An equation like
(\ref{selfcons}) would then appear as a constraint for possible
variations in the deformation of the self-consistent field. After
appropriate linearizations and after applying time dependent
perturbation theory in suitable manor one would end up in a form like
(\ref{chicoll}), provided that one is interested in the change of the
expectation value of the same operator $\hat F$ which specifies the
variation in the density. In general the response "function" would have
tensor character, of course.

In this context one may mention examples worked out for Landau Fermi
liquid theory in \cite{heispethravann},
\cite{kidhofpl}. Looking at 
general modes of the single particle density in phase space for the
associated response functions $\rfc{\om}$ forms similar to the one given
in (\ref{chicoll}) have been deduced. There it might only happen that
the "coupling constant" $k$ may depend on the wave vectors of the
various modes.

\subsection{Extension to finite excitations and large scale motion}

In the following we would like to concentrate on the example for which
a Hamiltonian $\ham{Q}$ is given, which depends parametrically on a
c-number $Q(t)$ representing the collective degree of freedom. It is
this example which is developed farthest theoretically and which still
can be handled numerically even for such complex situations as nuclear
fission \cite{ivhopaya}, \cite{yaivho}. As described in various papers
(see e.g. \cite{kidhofiva} and \cite{hofrep}, in particular), with such
a Hamiltonian it is possible to derive an intrinsically consistent
theory for collective motion of large scale provided there is a clear
separation of time scales.

Let $\tau_{\rm coll}$ be a relevant measure for a collective time scale
and $\tau$ the one for the remaining degrees of freedom, henceforth
called the intrinsic ones.  For time lapses $\delta t$ with $\tau
\le\delta t \ll \tau_{\rm coll}$ one may describe collective motion
within a (locally) harmonic approximation. This is achieved by
exploiting the expansion (\ref{hamexp}) and identifying simultaneously
the unperturbed density operator $\hat{\rho}_{\rm qs}^0$ as the one
corresponding to the Hamiltonian $\ham{Q_0}$. This density operator may
then either be a function of temperature or entropy depending which
distribution we want to use for parameterizing the equilibrium. For the
sake of simplicity, in this paper we will choose the canonical one
later on, but we may refer to \cite{hofrep} for a discussion of the
more general case. Thus we may write for the density $\hat{\rho}_{\rm
qs}^0= \hat{\rho}_{\rm qs}(Q_0,T_0)$, with the $Q_0$ and the $T_0$ to represent
those values which the "macroscopic" variables $Q,T$ have during the
time interval $\delta t$.  

As shown in \cite{kidhofiva} and \cite{hofrep} all the steps discussed
above for the genuinely harmonic case may then be taken over. We must
expect, of course, that all quantities will depend on the actual
quasi-static state of the system which is being specified by the pair
of variables $Q_0,T_0$. This is true in particular for the coupling
constant $k$ which can be written in the form
\bel{coupcon}
 -k^{-1} = 
   \left.{\partial^{2}E(Q,S_{0})\over \partial  Q^{2}}\right\vert_{Q_{0}}
   +\chi (0)\equiv C(0) + \chi (0) \, .
\ee
Here $E(Q,S_{0})$ is meant to represent the internal energy of the
system, with the entropy $S_0$ being understood to be calculated for
given $Q_0,T_0$. (It is possible, of course, to rewrite this formula in
terms of derivatives of the free energy). The $\chi (0)$ stands for the
static response and the $C(0)$ is introduced as a short hand notation
for the local static stiffness. To derive this relation, together with
the form (\ref{chicoll}) for the collective response function, one
needs to apply similar arguments as mentioned before but has to care
for possible changes in the quasi-static properties of the system. In
doing so one may be guided by the fact that within the harmonic
approximation (for the motion in $Q$) entropy can be assumed to be
constant (to leading order), as it will change with velocity only
quadratically.  Using this hypothesis, in formal sense the derivation
of the form of the collective response $\rfc{\om}$ follows identically
the one given above and which ended in (\ref{chicoll}), provided the
adiabatic susceptibility $\chi^{\rm ad}$ is identical to the static
response, $\chi^{\rm ad}=\chi(0)$; for a general discussion of the
relevance of this property  see \cite{hoivyanp} and  \cite{hofrep}.

The existence of equilibrium for the nucleonic degrees of freedom \
requires the presence of residual interactions. However, it would be
beyond the scope of any realistic theory to allow for both a detailed
treatment of such interactions as well as numerical applications, say for a
calculation of the transport coefficients which such a theory aims at.
It has been suggested, therefore, to acount for that by
dressing the particles and holes (of the deformed shell model) by a
self energy whose imaginary part is given by
\bel{imselfenomt}
\Gamma(\omega,T)=
    {1\over \Gamma_0}\;{(\hbar \omega - \mu)^2 + \pi^2 T^2 \over 
    1 +{1\over c^2}\left[(\hbar \omega - \mu)^2 + \pi^2 T^2 \right]}\,,
\ee
with $\mu$ being the chemical potential. It is this imaginary part
which finally will be responsible for the presence of macroscopic damping.
It would out of scope of the present paper to explain more details; 
instead, the reader is referred to \cite{hofrep}. In numerical
applications the following values have been used for the two parameters:
$\Gamma_0=33 MeV$ and $c=20 MeV$. 

With respect to the choice of $Q_0$, we should like to note that for
the general situation away from the potential minimum, the $Q_0$ in
(\ref{selfcons}) is to be replaced by the center $Q_m$ of the
oscillator approximating the "true" static energy $E(Q,S_0)$ 
in the neighborhood of the $Q_0$ of (\ref{hamexp}) (see
e.g.\cite{hofrep}).  Furthermore, we want to mention that the
discussion to come shall be restricted to situations of negative
coupling constants, which for the case of an instability will imply the
$\mid C(0) \mid$ to be smaller than $\chi(0)$, see below. Looking back
to the two body interaction introduced in (\ref{twobodham}) this is to
say that we like to restrict ourselves to effective forces being
attractive. For the nuclear case this is known to correspond to motion
with neutrons and protons in phase.

\subsection{Reduction to single modes}

The poles of $\rfc{\om}$ define possible excitations of the system, more
precisely those which involve dynamics in the quantity $\fmb$ and which
are caused by that perturbation  $\delta \hat H$ which is proportional
to the $\hat F$ itself. Considering all of them would in the end imply
strict non-Markovian behavior in $\fm{t}$. Eventually the strength
distribution is dominated by only one mode, in which case the
$\rfc{\om}$  may be replaced by the response function of the damped
oscillator 
\bel{resosc}
\chi^F_{\rm osc}(\om)= -{1\over M^F}{1\over \om^2+i\om{\Gamma}-\varpi^2}
\ee
with $ M^F \Gamma = \gamma^F$ and $M^F \varpi^2 = C^F$, where 
$\gamma^F,~M^F$ and $C^F$ stand for the coefficients of friction, inertia
and stiffness for the (local) motion in $\delta \fm{t}$, a feature
which is easily verified from (\ref{qext}). 
For several reasons it may be advantageous to introduce the coordinate  
$q(t)=Q(t)-Q_m$ for which the local equation of motion writes
\bel{eomaverq}
M \ddot q(t) + \gamma \dot q(t) +C q(t) = -q_{\rm ext}(t)
\ee
This transformation is readily performed considering (\ref{selfcons}),
with the $Q_0$ being replaced by $Q_m$, of course. It implies to work
with the $qq$-response function
\bel{qqresp}
k^2 \rfc{\om} = \chi_{qq}(\om) = \left. 
   {\delta q(\om) \over \delta (-q_{\rm ext}(\om)} \right\vert_{q_{\rm ext}=0} 
\ee
instead of the $\rfc{\om}$, with a similar relation leading from the
$\chi^F_{\rm osc}(\om)$ to the oscillator response $\chi_{\rm
osc}(\om)$ for q-motion. The latter is then specified by the transport
coefficients $\gamma =k^{-2}\gamma^F,~M=k^{-2}M^F$ and $C=k^{-2}C^F$. 

Whether or not such an ideal situation is actually met depends largely
on the value of the coupling constant $k$. In fair approximation one
may say it to be given whenever in (\ref{coupcon}) the static stiffness
$C(0)$ is small compared to the static response $\chi(0)$. In such a
case the strength of all the solutions of the secular equation $1/k +
\chi(\om) = 0 $ will concentrate in the low frequency domain. Therefore
it is important to realize that the static energy may depend strongly
on temperature. For nuclear physics, for instance, the $C(0)$ drops
dramatically above $k_BT= 2\;{\rm MeV}$ where the influence of shell
structure has diminished (see e.g.\cite{hoivyanp}).

In the more general case the strength distribution of the collective
modes, which is determined by the imaginary part $\chi^{\prime
\prime}_{qq} (\om)$ of the $qq$-response function, will spread over
several (or even many) modes. In this case the equation of motion for
$q(t)$, which may be obtained from
\bel{qeomnm}
\left(\chi_{qq}(\om)\right)^{-1}\, q(\om) = - q_{\rm ext}(\om) 
\ee
after a Fourier transform back to time, will contain terms being
non-local in time which cannot be expressed simply by a sum of
derivatives of $q(t)$ up to second order. If one still wants to stick
to differential form one needs to apply a further approximation. For
instance, one may simply expand the $\left(\chi_{qq}(\om)\right)^{-1}$
in (\ref{qeomnm}) to second order in $\om$ \cite{ivhopaya}. Another
possibility is found in "reducing" the $\chi_{qq}(\om)$ to the
$\chi_{\rm osc}(\om)$.  By this we mean to replace the former by the
latter in a certain range of frequencies. In practice this may be done
by fitting the imaginary part of $\chi_{\rm osc}(\om)$ to that peak in
$\chi^{\prime\prime}_{qq}(\om)$ the corresponding mode of which one
wants to treat. Notice please that both of these approximations require
that only frequencies  are relevant which lie in some limited regime.
In this sense the second one may be considered superior to the former.
In the past, various computations of transport coefficients have been
performed which lead to values being in fairly good agreement with
experimental experience, see \cite{yaivho}, \cite{ivhopaya}.

\subsection{Unstable modes}

The oscillator response is analytic in the stiffness $C$.  As we will
see below, the latter becomes negative whenever the static stiffness
$C(0)$ becomes negative, which happens in the region of potential
barriers where the system is unstable, in the sense that one of the
two fundamental solutions for average motion begins to grow
exponentially.  Such a situation can still be handled within the
locally harmonic approximation as long as the forces do not change too
rapidly with the coordinate. Recall please that at some given $Q_0$ the
harmonic solution is meant to portray the true one only within a
limited time interval $\delta t$ anyhow. This growth of the time
dependent solution reflects itself in the feature that for negative $C$
one of the poles of the collective response moves into the upper part
of the frequency plane. (Notice please that the instability we have in
mind still leaves both inertia as well as friction positive). This does
not cause any problems when dealing with the functional form of the
oscillator response as given by (\ref{resosc}), but it may well be so
for more general expressions.  For instance, one may want to write the
response function in terms of integral representations, like
\bel{rftodrfst}
\chi_{\mu\nu}(\om)=\int_{{\cal C}} {d\Om\over\pi} 
{\chi^{\prime\prime}_{\mu\nu}(\Om)\over\Om-\om}
\qquad {\rm for}\, \IM \,\om > 0\, .
\ee
In this case some caution is required in defining the contour $\cal C$.
For stable systems it is taken along the real axes with the poles lying
below. This choice is required by causality. If the same principle is
still to be valid for the case of an instability the contour  ${\cal C}$
must still be chosen to be above the one singular pole mentioned before.
Actually, besides this "retarded" function it may become desirable to
have an "advanced" response function as well, for which the contour
then has to be modified to lie below the poles which correspond to
motion backward in time. More details of these formal techniques are
described in Appendix A.

\subsection{Implications from sum rules}

The integral representation (\ref{rftodrfst}) may serve as a starting
point for deriving and exploiting sum rules, even in the case of
instabilities. Generalizing the common definition to
\bel{sumruldef}
S^{(n)}_{\mu\nu} = \int_{\cal C}{d\om\over\pi}\om^n\dissip_{\mu\nu}(\om)\,
,
\ee
the same relations are recovered for the unstable case as those known for
the stable one, if one only observes the appropriate choice of the contour.
For example, for $n=-1$ the value of the integral in (\ref{sumruldef})
gives the static response, $S^{(-1)}_{\mu\nu}=\chi_{\mu\nu}(\om=0)$, 
while for $n=1$ one may employ the usual arguments to obtain 
\bel{suruone}
S^{(1)}_{\mu\nu}= \int_{\cal C} {d\Om\over\pi} \Om\dissip_{\mu\nu}(\Om)=
             \lim_{\om\to\infty}\om\int_{\cal C}{d\Om\over\pi}
                    {\dissip_{\mu\nu}(\Om)\over 1-{\Om-i\eps\over\om}} =
             -\lim_{\om\to\infty} \om^2\chi_{\mu\nu}(\om)\, .
\ee
One way of getting the equality in the middle is to involve the
geometrical series for the $1/(1-(\Om-i\eps)/\om)$, which is
possible whenever the $S^{(n)}_{\mu\nu}$ remain finite for $n>0$.  The
latter restriction is not really necessary. One may simply replace the
$\Om$ in the first integral by $\Om =\om(1 + \Om/\om) = [1-(\Om/\om) -
(\Om/\om)^2/(1+\Om/\om)]^{-1}$ and observe that for $\om \to \infty$
the last term in the brackets gets less and less important. Then one
only needs to have the $S^{(0)}_{\mu\nu}$ vanish, which in our case is
no problem as later on we are going to apply (\ref{suruone}) only for
the diagonal case.  Please observe that for the oscillator response
(\ref{resosc}) the validity of the expression on the very right follows
directly from inspection.

The sum rules for $n=\mp 1$ are intimately related to two of the three
transport coefficients. This can be inferred from the values these two
sums take on for the oscillator response, namely
\bel{sumrulval}
S_{\rm osc}^{(-1)}= {1\over C}\qquad 
S_{\rm osc}^{(1)}={1\over M} \qquad {\rm with}\quad  S_{\rm osc}^{(0)} = 0
\, .
\ee
For later purpose we may note here that these results remain unchanged
when the oscillator is modified by introducing a frequency dependent
friction $\gamma=\gamma(\om)$.

As the oscillator response is derived from the original, microscopic
expression given by (\ref{chicoll}) together with (\ref{qqresp}), the
sum rules may be used to derive inequalities for the transport
coefficients. How these inequalities come about may best be seen
writing the full response function in terms of a pole expansion like
\bel{resoscsum}
 \chi_{qq}(\om) = -\sum_\alpha {1 \over 2M_\alpha {\cal E}_\alpha} 
   \left({1 \over \omega - \omega _\alpha^{+}} -
         {1 \over \omega -\omega_\alpha^{-}}\right)
\ee 
with $\omega_\alpha^{\pm}= \pm {\cal E}_\alpha - i \Gamma_\alpha/2$.
Then one may deduce results like those given in (\ref{sumrulval}) for
each term of the sum over $\alpha$. It is then evident that the value
of the  $S^{(\pm 1)}$ for each term cannot be larger than the
corresponding value of the total sum. 

Let us look at the inertia first. The form (\ref{chicoll}) implies
$\lim_{\om \to \infty} \om^2(\rfc{\om} - \chi(\om)) =0$. Therefore we
get 
\bel{sumqqin}
k^{-2}\,S^{(1)}_{qq}=\int^{+\infty}_{-\infty} \! {d\Om \over \pi}\, 
    \chi_{\rm coll}^{\prime\prime}(\Om)\, \Om =
    \int^{+\infty}_{-\infty} \! {d\Om \over \pi} 
    \, \chi^{\prime\prime}(\Om )\, \Om ={i\over \hbar}\langle [\hat  
    {\dot F},\hat F]\rangle 
\ee
and consequently
\bel{sumrulinert}
{1\over M} \le S^{(1)}_{qq}={ik^2\over \hbar}\langle [\hat  
    {\dot F},\hat F]\rangle \equiv {1\over m_0 } 
\ee
which means that the inertia for the one mode we have chosen to pick
cannot be smaller than the $m_0$ given by the microscopic value of the
energy weighted sum. In nuclear cases this $m_0$ can be
seen to correspond to the inertia one would get in the liquid drop
model for irrotational flow (see e.g. \cite{sije}).

Next let us look at the stiffness. Here the situation is slightly
different as the mode we like to look at explicitly may become
unstable.  However, assuming that all the other ones remain stable one
may simply repeat the arguments from before such that for
$S^{(-1)}\equiv\chi_{qq}(0)$ one gets
\bel{ineqstiff} 
{1\over C} \leq \chi_{qq}(0) \, = k{\chi (0) \over {1\over k}+\chi(0) }= 
{1\over C(0)}\,{\chi(0)\over \chi(0)+C(0)}  .
\ee
These expressions for $\chi_{qq}(0)$ are obtained after observing
(\ref{chicoll}), (\ref{qqresp}) and (\ref{coupcon}). The last equation
shows the $\chi_{qq}(0)$ to have the same sign as the static response
$C(0)$, from which observation it follows that $C$ will become negative
as soon as $C(0)$ is negative. In these arguments we have assumed the
static {\it intrinsic} response $\chi(0)$ to be positive, as there is
no reason to expect the intrinsic system to be unstable.  As mentioned
before, we want to stick to the case where this value is larger than
the absolute value of the static stiffness.  The inequality allows one
to prove that effective stiffness cannot be smaller than the static
one:
\bel{ineqstiffres}
C\geq C(0)
\ee
For stable modes this is immediately clear, for unstable ones it is
useful to make a detour and introduce the absolute values $C=-\mid C \mid$
and  $C(0)=-\mid C(0) \mid$, in which way one gets from (\ref{ineqstiff}) 
\bel{stiffboundfeinv} 
{1\over \mid C\mid } \geq {1\over \mid C(0)\mid }\,{1\over 1 - \mid
C(0)\mid /\chi(0)}  \geq {1\over \mid C(0)\mid }\; .
\ee
The inequality (\ref{ineqstiffres}) expresses in clear fashion that
in general the effective stiffness $C$ will be different from the
static one, $C(0)$. The reason for this difference is found in {\it
dynamical} effects contributing to the effective potential. It is
interesting to note that at an instability these effects actually {\it
lower the absolute value of the stiffness}, opposite to the case in
stable modes. In this sense the dynamics weakens the instability
by {\it de-}creasing the conservative force.

\subsection{Inferences for the high temperature limit}
Two body Hamiltonians of the type (\ref{twobodham}) are frequently used
in nuclear physics. Commonly, however, the notion of the coupling
constant $k$ is just taken in literal sense, namely to be a given
parameter which does neither change with deformation $Q$ nor with
temperature $T$ (see e.g. the review \cite{egidring}). This is an
essential difference to the situation described above for which $k$
does depend both on $Q$ and $T$, as given by (\ref{coupcon}). To see
the implications of the variation with $Q$ would require more extensive
studies, which are to be deferred to the future. In particular, it
would be necessary to extend the present treatment to the
multi-dimensional case, as given for genuine quadrupole vibrations, for
instance. Such a procedure is possible, in principle (see e.g.
\cite{hofrep}), but it involves efforts much beyond the present study.
For the $T$ dependence, on the other hand it is quite easy to see
important consequences, even at a qualitative level by simply looking at
the limit one would get for the collective mode when increasing the
nuclear excitation.

In the last decade or so, rather fancy procedures have been adapted from field
theoretical methods to study collective dynamics at finite $T$. Some of
them exploit the concept of small amplitude vibrations about some given
thermal state of the system, like RPA at finite temperature (see
e.g.\cite{egidring}) or the variants of the "Static Path Approximation"
such as the P(erturbed)SPA of \cite{attalha} or the C(orrelated)SPA of
\cite{roscan}. Also there separable two body interactions similar to
the one given in (\ref{twobodham}) are exploited (with the $k$ being
some given constant, it is understood). One of the striking results is
that with increasing $T$ the vibrational modes turn into the
unperturbed excitations of the nucleonic degrees of freedom, even for
those which at $T=0$ would show the typical collective behavior. In our
opinion, this feature contradicts basic concepts of Niels Bohr's model
of a compound nucleus. The latter is dominated by the effects of strong
(incoherent) two body correlations, or, to use a word which draws more
on analogies with transport theory, by the effects of two body
collisions.  There can be little doubt that the latter will become the
more important the larger the thermal excitation (c.f. the Ansatz for
the nucleon's self-energy suggested by (\ref{imselfenomt})). One may
then legimately ask, why it should be possible for the {\it
independent} particle motion configurations to dominate the dynamics at
large $T$?

At this stage it may be very worth while to recall the essence of the
Strutinsky renormalization of the total static energy. For small
excitations the variation of the latter is completely determined by
shell effects. However, what remains when they are gone is the liquid
drop energy. Consequently, the vibrations of the nucleus should behave
more like those of a "drop" of nuclear matter showing largely
"macroscopic" behavior, rather than that of independent particle motion. 
This is indeed what one gets from the theory described in the present 
paper (see e.g.\cite{hoivyanp} or \cite{hofrep}, with further references
given therein). As mentioned previously, with decreasing shell effects
the coupling constant drops to its macroscopic limit.  Consequently,
the modes get very soft. Even more important is a strong concentration
of strength in just {\it one broad collective mode}. The large width
associated with this has to be understood as clear signature of strong
damping, the latter feature being inherently related to the strong
correlations mentioned previously. Unfortunately, the strength
distribution of isoscalar modes has not yet been measured for finite
temperature. Such a measurement would be an ideal test of the
transition to the macroscopic limit \cite{hoivyanp}, \cite{hofrep}.
However, this transition reflects itself in transport coefficients for
slow large scale collective motion like fission, in particular in their
temperature dependence. Within the theory described above one finds
that damping increases with $T$, which is in at least qualitative
agreement with experimental findings, see \cite{yaivho} and
\cite{ivhopaya}. It should be noted that the relevant quantity is not
the friction coefficient itself, but combinations with those of
stiffness and inertia, depending on which quantities one looks at.
Whereas the $T$-dependence of the stiffness is well known and can be
deduced from calculations of the static energy, this is less clear for
the inertia. It has been shown first in \cite{hoyaje} that one of the
consequences of the $T$-dependent coupling constant $k$ of
(\ref{coupcon}) is a behavior of $M(T)$ as one would expect for a
transition to the macroscopic limit: At small excitations the $M$ shows
the typical features of the cranking model, but above the point where
shell effects have disappeared $M$ approaches values similar to those
of irrotational flow.

\section{Extension to dynamical fluctuations}

In the more general sense our treatment so far has not gone beyond a
mean field approximation. It is true, however, that the dynamics in the
basic quantities $\fm{t}$ or $q(t)$ may be damped. Microscopically this
damping implies that we need at least to consider effects which are not
contained in a pure Hartree (or Hartree-Fock) approximation. However,
it still suffices to stick to a description in terms of self-energies
in the single particle degrees of freedom \cite{hofrep} and the $q(t)$
still simply is a c-number.  Evidently the latter restriction has to be
given up whenever one is interested to describe processes where
fluctuations in the collective degrees of freedom become relevant. Take
the example mentioned before, the decay out of a potential minimum.
One knows from the pioneering work of Kramers \cite{kram} how in the
realm of classical physics the dynamics of fluctuations across the
barrier can be treated and how they influence the decay rate. In
\cite{hofthoming} it was shown how the quantal extension of this decay
rate can be obtained within a description based on transport equations.
Some of the ideas behind this derivation were borrowed from the method
developed in \cite{ingold}.  Unfortunately, also this method was again
based on functional integrals for which reason it can be applied only
to the schematic cases mentioned in the introduction.  In this section
we thus would like to proceed describing the basic steps needed for a
formulation in terms of transport equations, with an elaborate
discussion to come in the last two sections of the paper about the
essential ingrediences of this transport equation. By those we mean
these quantities which are intimately associated to fluctuations,
namely the coefficients which parameterize the diffusion process in
collective phase space.

\subsection{Quantization of the collective variables}

As the first step we need to render the collective degree of freedom to
become a fluctuating variable. Often people are tempted to do this
simply by adding to the Hamiltonian (\ref{hamexp}) a kinetic energy
term like ${\hat \Pi}^2 /(2 m )$, with some unknown mass parameter $m$
to represent the collective inertia and ${\hat \Pi}$ being the momentum
conjugate to the coordinate, which then is to be considered an operator
$\hat Q$.  It is not difficult to convince oneself that in this way
self-consistency would badly be lost; for a detailed discussion see
\cite{hofrep}. This feature cannot be cured by playing with the
potential part, which means to say by adding some function of $Q-Q_0$.
Rather one needs to introduce a momentum dependent coupling.

Starting from the effective Hamiltonian (\ref{twobodham}) in the
original particle space, this method allows one to deduce the following
Hamiltonian for the total system (see \cite{hoso}):
\bel{totham}
{\cal H} = \ham{Q_0} + k \hat \Pi {\hat{\dot F}} -
    {\beta \over k} (\hat Q - Q_0) \hat F + {\hat \Pi^2 \over 2 m_0 } + 
    {(2\beta + k) \over 2 k^2}(\hat Q - Q_0)^2
\ee
Comparing with (\ref{tothamdef}) one easily identifies the parts for
bare intrinsic and collective motion, as well as the coupling
between both, which here consists of the two terms being linear in
$\Pi$ and $Q-Q_0$.  In (\ref{totham}) the ${\hat{\dot F}} = i \;
[\ham{Q_0} , \hat F]$, the $ \hat \Pi$ stands for the canonical
momentum and $m_0$ is the inertia which by (\ref{sumrulinert}) was
introduced to represent the energy weighted sum, and which serves here
to define the unperturbed collective kinetic energy. The parameter
$\beta$ has to be specified only in connection to the collective
momentum, a point to which we will come to below. This $\beta$ drops
out of any equation of motion which only involves the collective
coordinate $Q$.  This feature is very similar to the properties we
found before for average dynamics (as expressed in $\fmb$ or in $q$)
which could be traced back to the response function  (\ref{chicoll}).
It should be no surprise that the same function appears again when the
dynamics for average motion in $q(t)$ is re-derived from Ehrenfest's
equations to (\ref{totham}). We may note in passing that in the
original version of the Bohm-Pines procedure, as it had been developed
for the electron gas in \cite{bopi}, no such additional parameter
$\beta$ was needed. The difference to our case lies in the fact that we
want to apply the method to an attractive two body interaction in
contrast to the repulsive one of the electron gas (c.f.\cite{hoso}).
 
The description in terms of the Hamiltonian (\ref{totham}) is not yet
complete. As it stands the ${\cal H}$ would have too many degrees of
freedom and in this sense could not be equivalent to the original 
two body Hamiltonian $\hat H^{(2)}$ of (\ref{twobodham}). Indeed,
the Bohm-Pines procedure necessarily involves a subsidiary condition
which for the present case reads
\bel{subcon}
k \hat F= {\hat Q}-Q_{0} 
\ee
It is nothing else but the "operator version" of the self-consistency
condition (\ref{selfcons}). This property and the very fact that it is
inherently incorporated into the form (\ref{totham}) of the Hamiltonian
${\cal H}$ makes it very plausible that the description of collective
motion on average is the same as before. The new feature is seen in the
fact that now we are in a position to treat fluctuations in the $\hat
q$ in the way similar to what the Hamiltonian $\hat H^{(2)}$ of
(\ref{twobodham}) allows one to do for the microscopic quantity $\hat
F$. This feature goes along with the fact that in (\ref{totham}) no two
body interaction appears anymore. Rather, its effects are hidden in the
terms involving collective coordinate and momentum.

With the Hamiltonian (\ref{totham}) one may proceed to derive effective
equations of motion for the propagation in collective phase space. In
this spirit the Nakajima-Zwanzig projection technique has been applied
in \cite{hosaoc} to get the transport equation for the Wigner function
of the collective density. However, to account for self-consistency not
only on the level of average dynamics, or on that of the mean field, a
non-perturbative method had to be developed for treating the integral
kernel. Again this was possible within the locally harmonic
approximation for which it suffices to know the propagators for the
consecutive time steps $\delta t$ with $\delta t\ll \tau_{\rm coll}$.
As for $\delta t \, \to\, 0$ this propagator starts at a definite point
in phase space, it was argued that a Gaussian approximation would do.
Thus one only needs to specify the time evolution of the first and
second moments. For this non-perturbative  Nakajima-Zwanzig equation it
was possible to prove that it is in accord with the (quantal)
fluctuation dissipation theorem (FDT) \cite{hosaoc}, which in full
glory requires to have non-Markovian equations of motion.  This feature
makes it possible to take advantage of the FDT for establishing the
approximate equations of motion, which for the first and second moments
may be taken to be of differential form.

\subsection{Harmonic approximation for first and second moments}

For average dynamics we have seen the differential equation of motion
to come up through the reduction of the "full" response function
(\ref{chicoll}) (or the one of (\ref{qqresp})) to that of the
oscillator given in (\ref{resosc}). Evidently, one would like to have
the equations of motion for the second moments to be of similar type.
Clearly, in addition to the coordinate they will involve the
collective velocity or the corresponding momentum as well. Let us first
introduce the latter on the level of first moments, i.e. for
$q_c(t)\equiv \langle {\hat q} \rangle_t$ and  $p_c(t)\equiv \langle
{\hat p} \rangle_t$. It is clear that for the harmonic case we are
looking at, one expects the following set:
\bel{diffeqqk}
M {d \over dt} q_c(t) = p_c(t) \qquad {d \over dt} p_c(t)
= -\gamma {d \over dt} q_c(t) -C q_c(t) \, ,
\ee
As the first equation tells one, the momentum appearing here is the
kinetic one, called $p$ henceforth. It is here where one may benefit from having the parameter
$\beta$ appearing in (\ref{totham}). As shown in \cite{hoso} and
\cite{hosaoc}, the choice $\beta + k =k^2 C$ allows one to replace the
canonical momentum $\Pi$ by the kinetic one not only for average motion
but also for the fluctuations around it. As demonstrated in
\cite{hosaoc} these fluctuations satisfy the following linear set of equations
\bel{smsigqq}
{d \over dt} \Sigma_{qq}(t) - {2\over M}\Sigma_{qp}(t) =0
\ee
\bel{smsigqp}
{d \over dt} \Sigma_{qp}(t)  
-{1\over M} \Sigma_{pp}(t)+ C\Sigma_{qq}(t)+{\gamma\over
M}\Sigma_{qp}(t)= D_{qp}
\ee
\bel{smsigpp}
{d \over dt} \Sigma_{pp}(t) +2C\Sigma_{qp}(t)+ {2\gamma\over
M}\Sigma_{pp}(t) = 2D_{pp}
\ee
The second moments are defined as $\Sigma_{qq}(t)= \langle {\hat q}^2
\rangle_t - q^2_c(t)$, and similarly for the combinations $\{q,p\}$ and
$\{p,p\}$. It is seen that the  structure of the homogeneous parts
follow from that of (\ref{diffeqqk}), for which reason the $q^2_c(t)$
etc. satisfy the set (\ref{smsigqq}) to (\ref{smsigpp}) with the
inhomogeneities $ D_{qp},\; D_{pp}$ put equal to zero. For any finite
value of these diffusion coefficients $ D_{qp}$ and $D_{pp}$ the
fluctuations will become finite in the course of time. This will be so
even for zero initial values, as it is the case when the solutions are to
represent the second moments of the propagators (see above and below).
So far the diffusion coefficients have not been specified. It is here
where we may benefit from the FDT.  As is easily verified, these
diffusion coefficients can be expressed by the stationary solutions of
(\ref{smsigqq}) to (\ref{smsigpp}).  Let us first take the conventional
case of a positive $C$ for which these stationary solutions are those
of equilibrium such that we may write:
\bel{diffcoef}
D_{pp}= {\gamma\over M}\Sigma_{pp}^{\rm eq}
\qquad
D_{qp}= C\Sigma_{qq}^{\rm eq} -{1\over M} \Sigma_{pp}^{\rm eq}
\ee 
In the next chapter we will exploit the (quantal) FDT to evaluate these
expressions. Afterwards we will turn to the case of an instability with
$C<0$ which will be handled by analytic continuation.

Often one encounters cases of strongly over-damped motion. It is not
difficult to convince one-selves (see \cite{hofrep} and
c.f.\cite{kram}) that in this case the set of equations reduces to
(\ref{smsigqq}) to (\ref{smsigpp})
\bel{smsigqqovd}
{d \over dt} \Sigma_{qq}(t) + 
    2 {C  \over \gamma} \Sigma_{qq}(t) =2 D_{qq}^{\rm ovd} 
\ee
with the diffusion coefficient $D_{qq}^{\rm ovd}=C\eqsigmaqq /\gamma$

It may be helpful to the reader to add some further comments on the
seemingly Markovian nature of our basic equations of motion. There can
be no doubt that approximations of this kind are necessary in one way
or other.  However, for average dynamics this step to differential form
does involve less stringent conditions than one would expect on usual
grounds. This feature is intimately related to the construction of the
oscillator response or the definition of the transport coefficients
$M,~\gamma$ and $C$. Look once more at (\ref{chicoll})  and suppose
that the microscopic response $\chi(\om)$ can be represented by a
Lorentzian, or its dissipative part, rather, which often may portray a
realistic situation quite well. For such a case the form (\ref{resosc})
follows without any additional approximation. On the other hand, it is
almost evident that the set of equations (\ref{smsigqq}) to
(\ref{smsigpp}) for the second moments can hardly be correct for all
times $t$ and in particular not for very small $t$. This set supposes
that we use it for large times, a feature already suggested by our way
of calculating the diffusion coefficients:  They are chosen in such a
way as to warrant that the set (\ref{smsigqq}) to (\ref{smsigpp})
describes correctly the relaxation to the proper equilibrium.

\subsection{Treatment of large scale motion}

The information contained in (\ref{diffeqqk}) to (\ref{smsigpp}) may
now be used to construct the time evolution in global sense. Two
possibilities come to one's mind. First of all and as indicated before,
one may take the first and second moments to define a Gaussian. The
latter may be constructed such as to represent motion within the time
lapse $\delta t = t_2 - t_1$ and which at $t=t_1$ starts from some
point $Q_1,~P_1$ in phase space. For a sufficiently short interval
$\delta t$ the widths $\Sigma_{\mu\nu}(t) $ may be assumed to remain
small such that this motion remains restricted to a narrow region in
phase space, thus justifying the linearization procedure we started
off. As this propagation can be performed for any point $Q_1,~P_1$, one
may in the end sum up all possibilities needed to construct the time
evolution of any initial distribution and even over time intervals of
macroscopically large size.  That this "propagator method" may be used
even to describe such complex problems as the decay out of a potential
minimum has first been demonstrated in \cite{scheuho}.

Numerically this method turns out to be time consuming as the
"summation over all possibilities" requires integration over the phase
space not only for the first step but for each following intermediate
step as well. It so turns out that it is simpler to translate it to
the picture of a Langevin force. There one follows individual
trajectories and constructs the final distribution by repeating this
latter procedure for a suitably large manifold of points representing
the initial distribution (in global sense). These trajectories are
constructed in such a way that for the elementary time step $\delta t =
t_2 - t_1$ one first moves along the average trajectory, say by solving
the set (\ref{diffeqqk}), to find the point representing the actual
position at time $t_2$ by Monte Carlo techniques. This sampling is to
be performed by taking into account the size the fluctuation attains at
$t=t_2$ according to the fluctuating force, which in turn may be
determined from the information contained in the diffusion coefficient.

\section{Diffusion coefficients for stable modes}
\label{undistrancoeff}

To evaluate the diffusion coefficients from (\ref{diffcoef}) we need to
calculate the values the fluctuations would take on if the
(collective) system were in equilibrium locally. The fluctuation
dissipation theorem implies
\bel{eqaverqq}
\Sigma^{\rm eq}_{\nu\mu}= \hbar \int {d\om\over 2\pi} 
\coth \left( {\hbar\om\over 2T}\right) 
\chi^{\prime\prime}_{\nu\mu}(\om)\, .
\ee
Here the indices $\nu,\mu$ are supposed to represent $q$ and $p$. 
Exploiting the relations
$\chi^{\prime\prime}_{qp}(\om)=iM\om\chi^{\prime\prime}_{qq}(\om)$  and
$\chi^{\prime\prime}_{pp}(\om)=(M\om)^2\chi^{\prime\prime}_{qq}(\om)$,
all fluctuations can be expressed by the one response function $\chi_{qq}$.  
The different behavior of $q$ and $p$ under time reversal causes the
cross term to vanish, i.e. $\Sigma^{\rm eq}_{qp}=0$.  Formally this is
reflected by $\chi^{\prime\prime}_{qp}$ being even in $\om$. It is for
this reason, that no diffusion coefficient appears in (\ref{smsigqq}).

Actually, the values given by (\ref{eqaverqq}) are not yet suitable for
our purpose. It was mentioned before that the equations (\ref{smsigqq})
to (\ref{smsigpp}) should be consistent with those for average motion
given in (\ref{diffeqqk}). As the latter are derived from the
oscillator response we are forced to take the latter also when
evaluating the fluctuations from an expression like (\ref{eqaverqq}).
In other words, the diffusion coefficients ought to be calculated from
relations like (\ref{diffcoef}) but where the $\Sigma^{\rm
eq}_{\mu\nu}$ are replaced by those for the oscillator $\Sigma^{\rm
osc}_{\mu\nu}$,
\bel{dkinqqpp}
D_{pp} = {\gamma \over M}\Sigma^{\rm osc}_{pp}  
\qquad {\rm and} \qquad  
D_{qp} = C \Sigma^{\rm osc}_{qq} - {1\over M}\Sigma^{\rm osc}_{pp}  \, ,
\ee
and the equilibrium fluctuations are finally to be calculated from
\bel{eqaverqqosc}
\Sigma^{\rm osc}_{qq}= \hbar \int {d\om\over 2\pi} 
\coth \left( {\hbar\om\over 2T}\right) 
\chi^{\prime\prime}_{\rm osc}(\om)\,
\ee
and
\bel{eqaverpposc}
\Sigma^{\rm osc}_{pp}= M^2 \hbar \int {d\om\over 2\pi} 
\coth \left( {\hbar\om\over 2T}\right)\, \om^2\,
\chi^{\prime\prime}_{\rm osc}(\om)\, .
\ee
In doing so we keep the $\Sigma^{\rm eq}_{qp}=0$ {\it untouched}, as it
results from a fundamental symmetry. 

\subsection{Limiting cases}

Before turning to the general case let us  look at the limits of high
temperature and of small damping. The former is established if the
argument of the hyperbolic cotangent is small, i.e. $\hbar\om/(2T)<<1$,
for all frequencies which contribute to the integrals. Quite generally,
one finds: $\Sigma^{\rm eq}_{\nu\mu}= T \chi_{\nu\mu}(\om=0)$\, which
for the oscillator leads to the classical equipartition theorem:
\bel{equipart}
{\langle\;\hat P^2\;\rangle_{eq}\over 2M} = {T\over 2}=
{C\langle\;\hat q^2\;\rangle_{eq}\over 2}.
\ee
As a consequence,  for the diffusion coefficient one obtains the
classic Einstein relation 
\bel{einstein}
D_{pp} = \gamma T \qquad {\rm with} \qquad D_{qp} = 0
\ee
Notice that for this result no assumption was necessary about the
strength of the friction coefficient, besides the fact that $\gamma $
will influence the range of frequencies which must be considered small
compared to $T$.

In the limit of zero damping one finds a similar result, only that the
temperature $T$ has to be replaced by an effective one, namely
\bel{effT}
T^*(\varpi)={\hbar\varpi\over 2}\coth\left({\hbar\varpi\over 2T}\right)
\ee
with $\varpi^2=|C|/M$. 
This is easily verified letting the friction $\gamma $ in the oscillator 
response (\ref{resosc}) go to zero. 
Then the dissipative part of the response function can be written as
\bel{disoscsd}
\lim_{\gamma\to 0}{\chi^{\prime\prime }_{\rm osc}(\om )} = 
{\pi\over 2M\varpi} 
\left( \delta(\om-\varpi)- \delta(\om+\varpi) \right) 
\ee
The resulting version of the generalized Einstein relation, i.e.
$D_{pp}=\gamma T^\ast$,  has been used in \cite{hogngo}.  It allows for
a nice and simple evaluation of the correction terms to the classical
relation, namely $D_{pp} \approx \gamma T(1+ (1/3)(\varpi/2T)^2$.
However, as we will demonstrate below, the  form $D_{pp}=\gamma T^\ast$
approximates the correct formula only for extremely small friction.

\subsection{General case}
 
To evaluate the integral in (\ref{eqaverqq}) one commonly expands
the hyperbolic cotangent into the uniformly convergent series
\bel{seriecoth}
\coth\left({\hbar\om\over 2T}\right)=
{2T\over\hbar}\sum_{n=-\infty}^{n=\infty} 
{1\over \om-in\Theta}\qquad\qquad \Theta={2\pi T\over\hbar}
\ee
and applies a general relation (see (\ref{rftodrf})) which allows one
to evaluate the response function through its spectral density, the
dissipative part;  such expressions are briefly discussed in Appendix
A. For the terms with positive $n$, the integral leads to the retarded
response function calculated at the corresponding pole of the $\coth$,
namely
\bel{contcaus}
\int_{{\cal C}} {d\om\over\pi} 
{\chi^{\prime\prime}_{\nu\mu}(\om)\over \om-i|n|\Theta}=
\chi^{\rm R}_{\nu\mu}(i|n|\Theta)
\ee
For negative $n$, one may use a relation between retarded and advanced
response to find
\bel{contacaus}
\int_{{\cal C}} {d\om\over\pi} 
{\chi^{\prime\prime}_{\nu\mu}(\om)\over \om+i|n|\Theta}=
\Big(\chi^{\rm R}_{\mu\nu}(i|n|\Theta)\Big)^\ast
\ee
Notice please that in the classical limit it is only the term with
$n=0$ which survives and which leads to $T\chi_{\nu\mu}(\om=0)$. 
The quantum corrections are caused by those contributions to
(\ref{seriecoth}) having a finite $n$.

A straightforward calculation then shows the diagonal 
case 
elements
$\Sigma^{\rm eq}_{\mu\mu}$ to be given by 
\bel{eqqqserie}
\Sigma^{\rm eq}_{\mu\mu} = T\left( \chi_{\mu\mu}(\om=0)+
     2\sum_{n=1}^{\infty} \chi_{\mu\mu}(in\Theta)   \right)
\ee
In this expression we have used the fact that the response function 
$\chi_{\mu\mu}$ is real along the imaginary axis.

To evaluate the diffusion coefficients we need to apply
(\ref{eqqqserie}) to the oscillator response. Unfortunately,  the
resulting series diverges logarithmically for the momentum fluctuations.
Actually, such a behavior can better be read off from the integral
(\ref{eqaverpposc}) after recognizing that the dissipative part of
(\ref{resosc}) decreases with $\om$ only like $\om^{-3}$.  To cure this
problem one needs to regularize the integral.  In \cite{hosaoc} the
latter was simply cut off at a certain upper limit. Here we would like
to apply the somewhat more elegant method of the so called Drude
regularization (see e.g. \cite{graschring}). It implies replacing in
(\ref{resosc}) the friction coefficient by a function of the form
\bel{drudegom}
\gamma(\om)= {\gamma\over 1-i{\om\over\omd}} \, .
\ee
Properties of the resulting modified oscillator response function are
discussed in Appendix \ref{sundrudereg}. In (\ref{drudegom}) the
additional parameter $\omd$ appears, the appropriate choice of which
depends on general properties of the system. In the present work we
shall assume the time scale $1/\omd$ to be well separated from that of
collective motion and will examine how in a physically reasonable
regime this parameter influences the diffusion coefficients.

For the $\Sigma^{\rm osc}_{pp}$ one finds the following expression:
\bel{eqppserie}
\Sigma^{\rm osc}_{pp}(\omd) = T M\left(1 + 
     2\sum_{n=1}^{\infty} \left[C+ n\Theta \gamma(i n\Theta) \right]
    \chi_{\rm osc}(i n\Theta) \right),     
\ee
A common procedure to evaluate further both (\ref{eqqqserie}) and
(\ref{eqppserie}) is to make use of the digamma function $\Psi(z)=
{d\over dz}\ln\Gamma(z)$.  It allows one to sum up series over products
of pole terms, for instance in the form (see e.g.
\cite{abramsteg,hansen})
\bel{difftwopsi}
\sum_{n=1}^{\infty} {1\over (n+y)(n+z)}={\Psi(1+y)-\Psi(1+z)\over y-z} 
\ee
The oscillator response with the constant $\gamma$ replaced by the one
of (\ref{drudegom}) has three poles $\om^{(i)}$ (see Appendix
\ref{sundrudereg}). Thus one may rewrite it as a sum of three 
terms where each one factorizes like the left hand side of
(\ref{difftwopsi}). After some manipulations one then ends up with
\bel{sigqqsym}
C \Sigma^{\rm osc}_{qq}(\omd) = T\Bigg(1+{C\over M \pi T}
  \sum_{j=1}^3 {\omd-i\pol{j}\over(\pol{j}-\pol{j+1})(\pol{j}-\pol{j+2})}
   \;\Psi\left(1+i{\hbar\pol{j}\over 2\pi T}\right)\Bigg)
\ee
and
\bel{sigppsym}
{1\over M}\Sigma^{\rm osc}_{pp} (\omd)=  C \Sigma^{\rm osc}_{qq} (\omd) +
 {\gamma\hbar\omd\over M\pi}
  \sum_{j=1}^3 {-i\pol{j}\over (\pol{j}-\pol{j+1})(\pol{j}-\pol{j+2})}
   \;\Psi\left(1+i{\hbar\pol{j}\over 2\pi T}\right)
\ee
where the convention $\pol{j+3} \equiv \pol{j}$ is used.  In
(\ref{sigqqsym}) also the $qq$-fluctuation has been evaluated with the
Drude regularization, mainly to allow for a clear comparison to the
other case where this regularization is really necessary.  It is clear
from (\ref{drudegom}) that for large $\omd$ the result (\ref{sigqqsym})
will turn into the one calculated for constant friction.  How rapidly
this happens depends on the values of the parameters involved. This is
demonstrated in Fig.1 a by numerical calculation.  
\begin{figure}[htb]
\centerline{
{\epsfysize=12.5cm 
\epsffile[41 63 539 754]{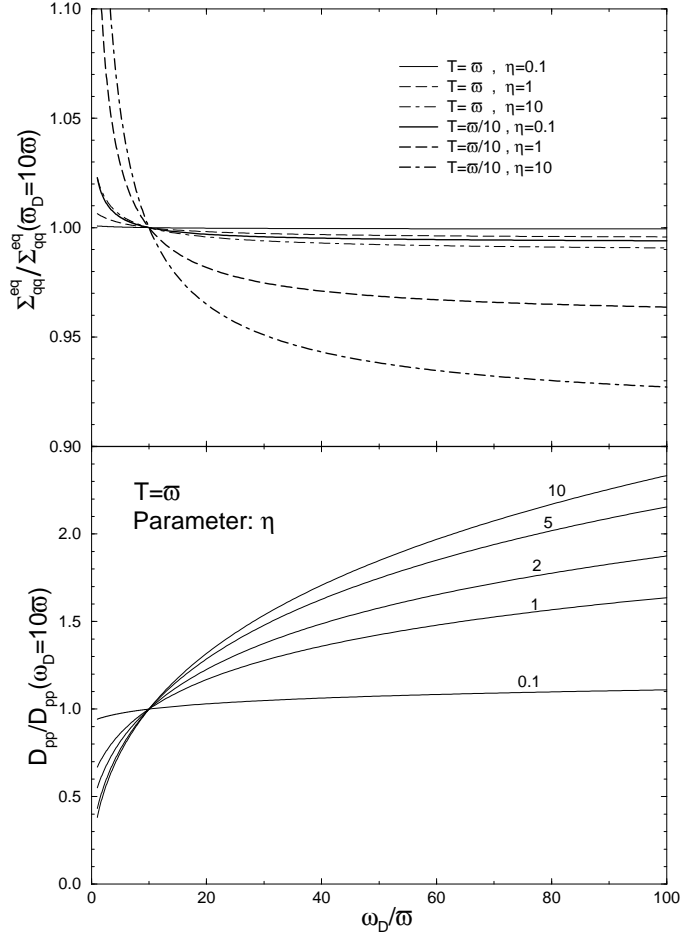}}}
\caption{\label{figomd}
Equilibrium fluctuations in $q$ (upper panel) and diagonal diffusion 
coefficient (lower panel) as functions of the cut off frequency for 
various values of damping rate $\eta$ and temperature $T$. (The
calculations are performed in units with $k_B =1$ and $\hbar =1$).
}
\end{figure}
It is seen that
large values of $\omd$  are needed for small values of temperature and
large damping before the asymptotic limit is really reached. It may be
seen as well that even for comparatively small $\omd$ the difference to
this asymptotic limit is quite small. (It should be clear from above
that $\omd$ must not be chosen smaller than the typical frequencies of
the original oscillator.) Here and in the following it turns out
convenient to measure frequency in units of $\varpi=\sqrt{|C|/M}$ and
the degree of damping by the dimension-less ratio
$\eta=\gamma/(2M\varpi)$.

Different to the $qq$-fluctuations, those of the momentum would diverge
for large $\omd$. Therefore, one must expect them to exhibit a stronger
dependence on the Drude frequency $\omd$ than seen in the
$qq-$fluctuations.  This is demonstrated in  Fig.1.b.  Indeed, for
large $\eta$ this dependence is quite sensible.  We may notice that the
diffusion coefficient increases with $\omd$, although its classical
limit is independent of $\omd$. In this sense it can be said the
quantum corrections to increase with $\omd$, too.  The standard values
chosen for the figures presented here, namely $\omd \approx 10\, \om$,
are probably very reasonable. Fortunately, in this region the
dependence of the diffusion coefficient on $\omd$ is not very strong,
the value of $D_{pp}$ changing by 30\% if $\omd$ is increased by a
factor 2.

Let us turn now to the dependence of the diffusion coefficients on
temperature $T$ and on the effective damping rate $\eta$.  Although the
transport coefficients for average motion, $M, \gamma$ and $C$, must be
expected to vary with these parameters as well, this feature will be
neglected here. All computations will be done in Drude regularization.
If not stated otherwise, the value of $\omd$ will be chosen such that
the ratio $\zeta=\varpi / \omd$ equals $0.1$.  To simplify notation we
want to use the same symbol as if we would use the mere oscillator,
which is to say that in expressions like the ones on the left hand side
of (\ref{sigqqsym}) and (\ref{sigppsym}) the argument $\omd$ will be
left out. We shall also omit the index $\rm osc$ at times.

Let us begin looking at the equilibrium fluctuations (see also
\cite{graschring}), first in the coordinate $q$. Rather than showing
the $\oscsigmaqq$ itself, we plot $C\oscsigmaqq$, normalized to its
value as given by the quantal ground state (Fig.2.a). Fig.2.a shows the
temperature dependence for different values of $\eta$.  
\begin{figure}[htb]
\centerline{
{\epsfysize=12.5cm \epsfxsize=10.5cm 
\epsffile[120 129 487 628]{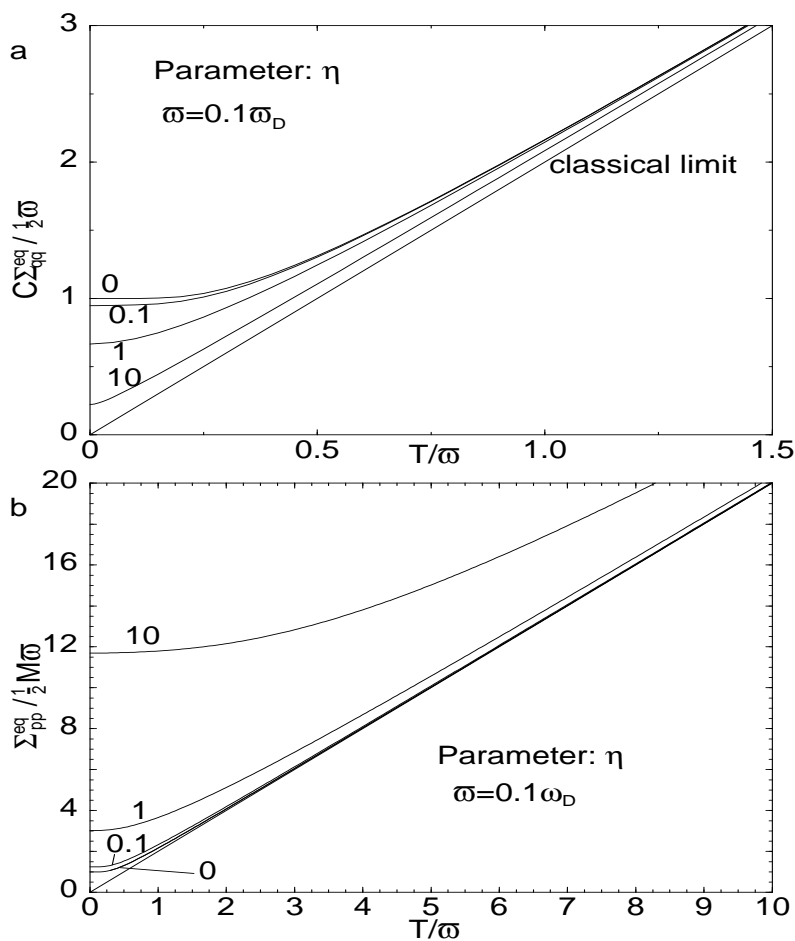}}}
\caption{\label{figsigt}
Equilibrium fluctuations in the coordinate (upper panel)
and in the momentum (lower panel) as functions of temperature
normalized to the frequency $\varpi$ of the oscillator, for different
values of the damping rate $\eta$.}
\end{figure}
One finds the
quantum effects to disappear both with increasing $T$ and $\eta$. The
first effect is expected, although it may be somewhat of a surprise how
quickly the high temperature limit is reached. The second effect causes
the $\eqsigmaqq$ to reach this limit the faster the larger the damping.
While behaving similarly with increasing $T$, the fluctuations in
momentum differ in their dependence on $\eta$: They increase with
increasing damping.  This is demonstrated in  Fig.2.b. Here
$\oscsigmapp/0.5 M \varpi$ (or equivalently $D_{pp} /0.5 \gamma
\varpi$) is plotted versus temperature for different $\eta$.  This
figure clearly shows how much the quantum effects are enhanced by
friction.  Take the regime of $T\leq \varpi$ where the change with $T$
is quite small. For large $\eta$ the fluctuations exceed several times
the values they have for undamped motion.

Next we study the diffusion coefficients. They are plotted in Fig.3~ by
the dashed lines as function of $\eta$ for different temperatures, the
latter being normalized to $\varpi$.  
\begin{figure}[htb]
\centerline{
{\epsfysize=11.5cm \epsfxsize=10.cm 
\epsffile[114 133 485 602]{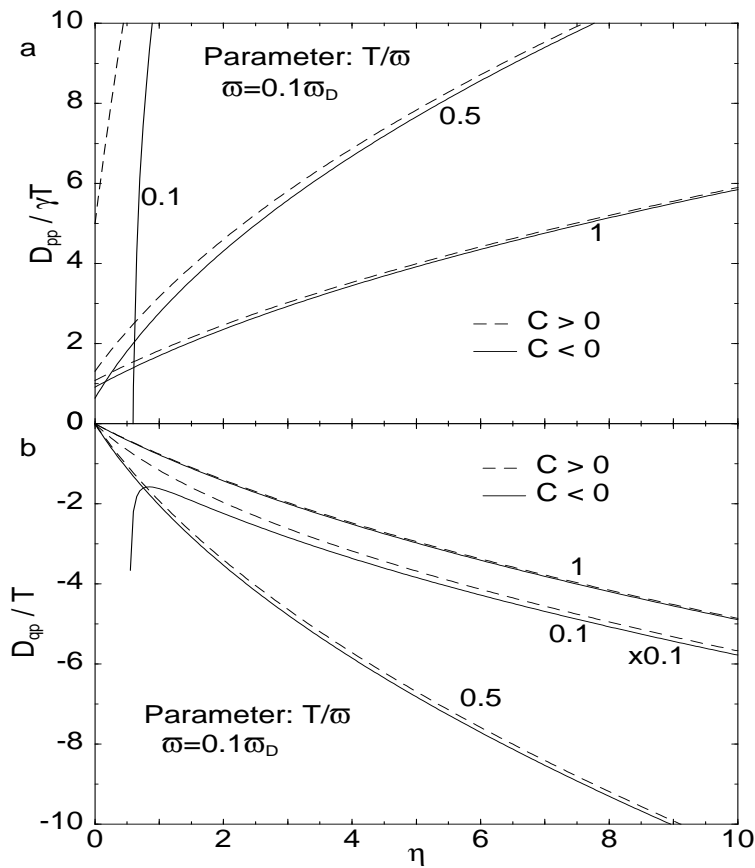}}}
\caption{\label{figdeta}
Diagonal (upper panel) and off diagonal (lower panel) diffusion
coefficient for stable (dashed) and unstable (fully drawn) modes, as
functions of the damping rate $\eta$, for three different temperatures.
In the lower panel the curve for $T=0.1\varpi$ shows $0.1 D_{qp}/T$, as
indicated by the symbol $\times 0.1$. 
}
\end{figure}
In Fig.3.a we show the diagonal
one, divided by its high temperature limit, $D_{pp}/\gamma T$;
Fig.3.b~represents the non-diagonal one, divided by temperature. Both
figures demonstrate that for the diffusion coefficients quantum effects
increase with damping, and that they may become quite large. Take the
non-diagonal element plotted in Fig.3.b~and recall its definition
according to (\ref{dkinqqpp}). It measures the difference between $C
\Sigma^{\rm osc}_{qq}$ and ${1\over M}\Sigma^{\rm osc}_{pp}$, which
vanishes in the high temperature limit.  But Fig.3.b~ demonstrates that
for small $T$ this difference, and thus $D_{qp}$, reaches values which
are several times this unit.  Moreover, these figures show that the
"limit of zero friction" discussed previously must be taken literally:
It is obtained for such small values of friction only that it is
practically never realized in nuclear physics. Look for instance at
Fig.3.b: Already for $\eta>0.5$ the $D_{qp}$ is seen to be at least of
order $T$ even for $T
\approx \varpi $, whereas in the strict limit of zero friction $D_{qp}$
vanishes identically.

It has been mentioned that in the computations of \cite{hosaoc}
regularization of the appropriate integral had been achieved by
introducing an upper limit in the integral itself, rather than to
change the integrand applying the Drude regularization.  The two
methods are compared with each other in Fig.4. They are seen to lead to
the same dependence on $\eta$ but may require an optimization of the
cut-off parameters, if the comparison is to become more quantitative.
\begin{figure}[htb]
\centerline{\rotate[r]
{\epsfysize=10.5cm \epsfxsize=8.cm 
\epsffile[95 93 549 671]{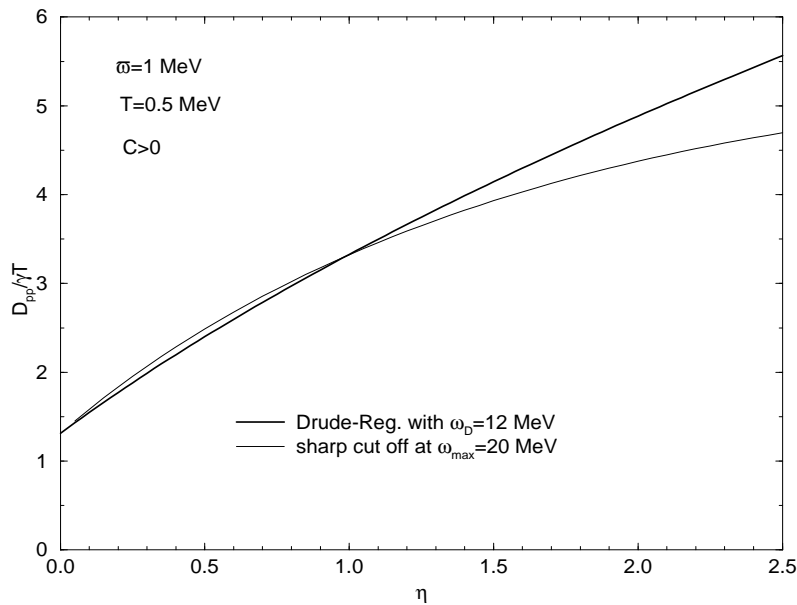}}}
\caption{\label{figregcom}
Comparison of two regularization schemes: Sharp cut off
(thin line) and smooth cut off through Drude regularization. The values
of the cut off frequencies, $\om_{\rm max}$ and $\om_{\rm D}$
respectively, were choosen in such a way that in both cases the
diffusion coefficient attains about the same value. 
}
\end{figure}

\section{Diffusion coefficients for unstable modes}
\label{sunquantdiffinst}
The principle which allows one to deduce diffusion coefficients from
the fluctuation dissipation theorem even for unstable modes has been
suggested in \cite{hofing}, see also \cite{hofthoming}. In the
following we like to explore this possibility in more detail with the
help of formulas obtained by adapting linear response theory to
instabilities.  All we need to do is to generalize expressions such as
(\ref{sigqqsym}) and (\ref{sigppsym}) to the case of negative
stiffness.

The basic idea is found in realizing that all the expressions derived
and discussed in the previous sections are analytic expressions in the
transport coefficients, and thus also in the stiffness coefficient $C$.
This is true in particular for the general expressions for the
diffusion coefficients as given by (\ref{sigqqsym}) and
(\ref{sigppsym}), together with (\ref{dkinqqpp}).  In Fig.5~we present
a computation of three typical quantities, namely $C\Sigma^{\rm
eq}_{qq}$, $D_{pp}$ and $D_{qp}$, all three normalized to temperature,
as function of the stiffness parameter $C$ (divided by the inertia $M$
and relative to $(\gamma/M)^2$). 
\begin{figure}[htb]
\centerline{\rotate[r]{\epsfysize=12.5cm \epsfxsize=8.cm 
\epsffile[104 58 549 669]{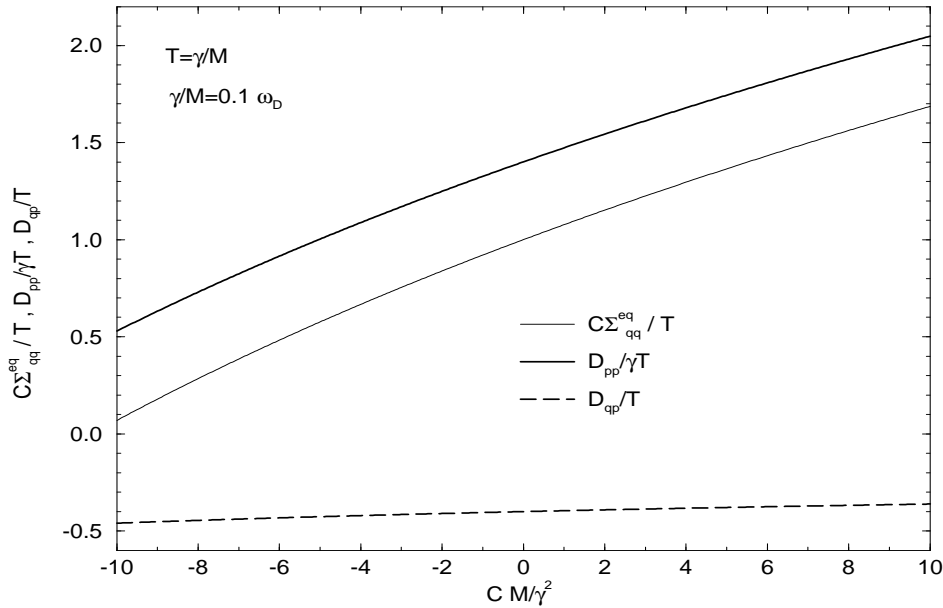}}}
\caption{\label{figdsigc}
Diffusion coefficients (diagonal: thick fully drawn line,
off diagonal: dashed line) and analytic continuation of the equilibrium
fluctuations in $q$ (thin full line) as functions of the stiffness $C$,
for fixed values of $M$ and $\gamma$. 
}
\end{figure}
The result demonstrates the
smoothness, or analyticity, across the point $C=0$, which actually
corresponds to the inflection point $C(0)=0$ of the static energy. This
feature is quite interesting, for several reasons. First of all one
realizes that the fluctuation in the coordinate diverges when $C$
approaches zero from above: $\Sigma^{\rm eq}_{qq} \to +\infty $ for
$C\to 0^+$. Secondly, across this point this "fluctuation" gets even
{\it negative}.  Fortunately, for the equations of motion it is not
this "hypothetical" {\it static fluctuation} which matters but the
diffusion coefficients.  The figure nicely demonstrates that all of
{\it them} remain well behaved at this point implying that the same
will be true for the {\it dynamical} fluctuations!

On the other hand, Fig.5~also demonstrates that there are limitations
in applying the analytical continuation to get meaningful diffusion
coefficients. It should be clear that the diagonal one ought to be
positive. However, depending on the parameters involved, like the
transport coefficients for average motion or temperature, the $D_{pp}$
may become zero or even negative.  This feature can be seen more
clearly in Fig.6~where we plot the analytical continuations for
$\Sigma^{\rm eq}_{qq}$ and $\Sigma^{\rm eq}_{pp}$ as function of
temperature divided by the bare frequency, i.e.  of $T/\varpi$. 
\begin{figure}[htb]
\centerline{\rotate[r]{\epsfysize=12.5cm \epsfxsize=9.5cm 
\epsffile[95 73 549 671]{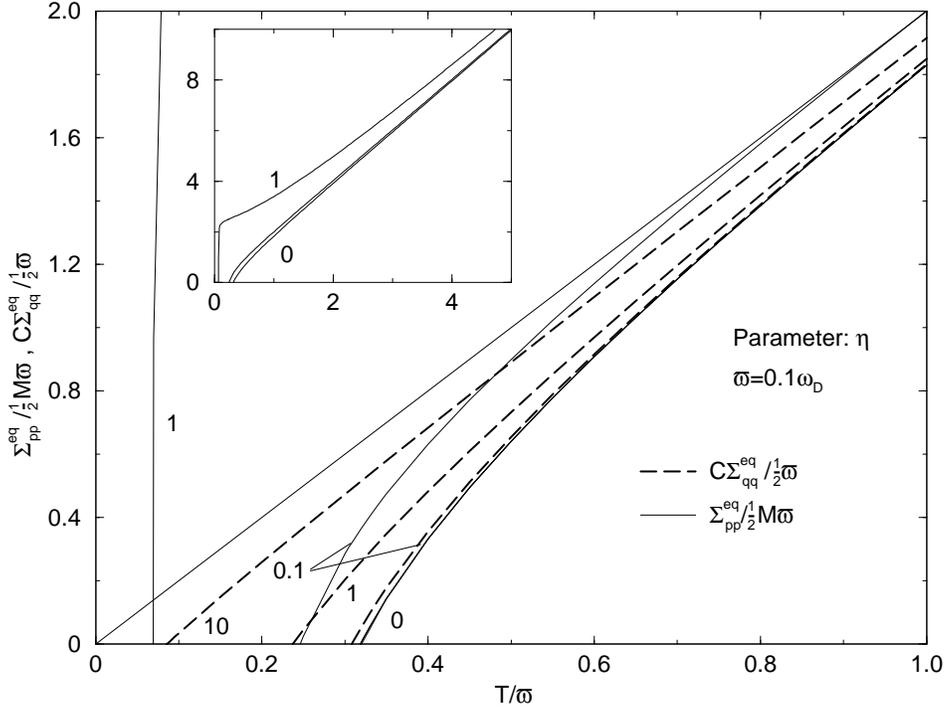}}}
\caption{\label{figsigtin}
Analytic continuations of the (diagonal) equilibrium
fluctuations to the case of negative stiffness (dashed line for the
coordinate, fully drawn lines for the momentum), as functions of
temperature for different values of friction. The insert shows the
momentum fluctuations for a larger range of the temperature and the
same values of $\eta$.
}
\end{figure}
The two
quantities are normalized such that in the high temperature limit they
approach the same value, namely $2T/\varpi$.  Actually, the factor two
here is a relict of the discussion of the stable case presented in
Fig.2, where the same quantities were plotted, and where this
normalization was convenient at $T=0$. Since the curves of Fig.6~just
represent the analytical continuation of those in Fig.2, it is not
surprising that the quantum effects behave in the same way as for
stable modes when studied as functions of temperature and friction.

At this stage it may be worth while to refer the reader to the two easy
examples discussed in the previous chapter, namely the limit of high
temperature on one hand and the case of zero damping on the other one.
Indeed, eq.(\ref{einstein}) tells one that the relevant diffusion
coefficient $D_{pp}$ remains positive for all $T$, but from
eq.(\ref{equipart}) on sees the  $\Sigma^{\rm eq}_{qq}$ to change sign
at the point of inflection. On the other hand, the $T^*$ gets negative
below a critical temperature of $T_c=\hbar \varpi /\pi$.  The figure
Fig.6~shows such features to hold true in the general case with the
following two important modifications: a) The $T_c$ is shifted to lower
values when $\eta$ is increased; b) this shift is bigger for the
momentum fluctuations. At larger values of $\eta$ the latter even show
a different behavior, in the sense that they become larger than the
value given by the high temperature limit, which for larger $T$ they
approach from above.

The non-diagonal diffusion coefficient $D_{qp}$ is plotted in Fig.7~as
function of temperature, both for positive and negative stiffnesses,
and in Fig.3.b~by the fully drawn lines as function of friction. 
\begin{figure}[htb]
\centerline{\rotate[r]{\epsfysize=12.5cm \epsfxsize=8.5cm 
\epsffile[103 60 549 671]{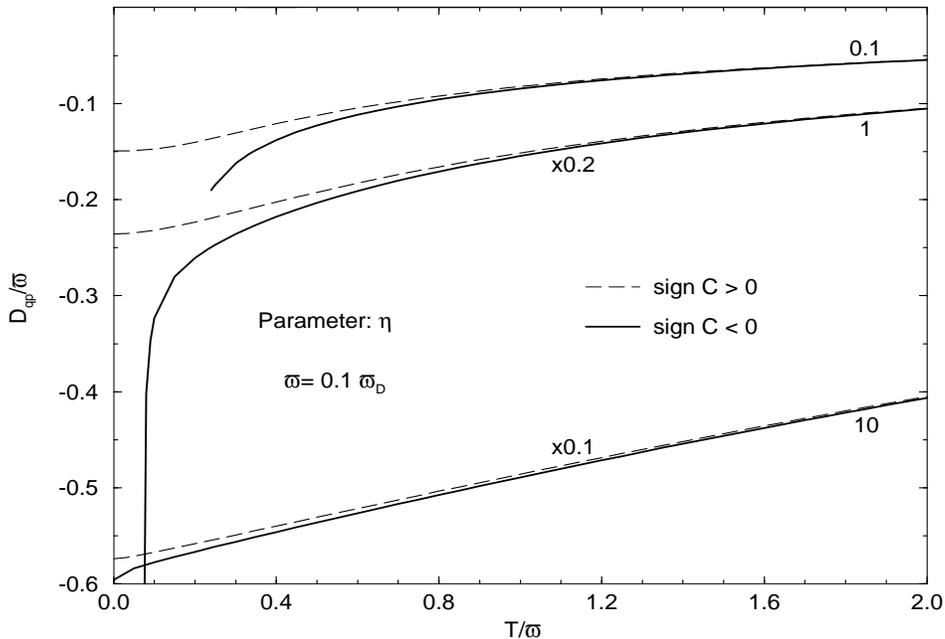}}}
\caption{\label{figdqpt}
Off diagonal diffusion coefficient as function of
temperature for three values of friction, for a stable (dashed lines)
and an unstable (fully drawn lines) system. The curves for $\eta=1,10$
show $D_{qp}$ multiplied by 0.2 and 0.1, respectively.
}
\end{figure}
The
values of $D_{qp}$ for positive and negative stiffness differ from each
other only for small temperatures and small friction, and practically
agree with each other already for temperatures larger than about $0.5
\hbar\varpi$ and friction $\eta > 2$. A similar behavior is true for
the diagonal element, as may be inferred comparing fig.~3.a and Fig.6.
At very small $T/\varpi$ the  $D_{qp}$ shows an anomalous decrease with
$T/\varpi$ (see Fig.7~) or with $\eta$ (see Fig.3.b~). This actually
happens in a regime below $T_c$, the reason of which will be discussed
on below.

Finally, we would like to come back once more to the fluctuation
dissipation theorem in its general form. We recall that it
was this theorem which allowed us to determine the diffusion
coefficients in terms of the response function of average motion.
Furthermore, these are exactly the expressions, taken as analytical
functions of the stiffness $C$, which we were able to generalize to
unstable modes by analytical continuation from a positive stiffness to
a negative one. The construction of the special contour ${\cal C}$ now
allows us to perform these steps in a more direct fashion, provided one
more constraint on the contour is considered. That is to say that at
instabilities we may express the "equilibrium fluctuations" by the
following integral:
\bel{bcoreqinv}
\Sigma_{\mu \nu}^{\rm eq}=\hbar \int_{\cal C} \, {{\rm d}\om \over 2\pi }\,
  \coth \left( {\om \over 2T}\right) \, \chi^{\prime\prime}_{\mu\nu} (\om)
\ee
In this integrand there appears another factor having poles in the
complex plane, the $\coth $ which diverges at the Matsubara frequencies
$ \om_{\rm M}^n =\pm  n {2\pi T}/\hbar $. As in the stable case, the
contour ${\cal C}$ has to cross the imaginary axis in between those which
lie closest to the real axis \cite{kiderlendoc}, namely $\pm  {2\pi T/
\hbar}$. This condition actually puts a lower limit on the range of
temperatures for which such a construction may work, namely $T > T_0
\equiv \hbar \vert\om_+\vert/(2\pi)$, with $\om_+$  being the frequency
of the unstable mode (see Appendix B, in particular
eqs.(\ref{carsolper}) and (\ref{thirdsol})  and the text below them).
This follows from the constraint discussed in the context of
eq.(\ref{rftodrfst}) according to which the contour ${\cal C}$ has to lie
{\it above} the unstable pole. Actually, this $T_0$ coincides with the
so called "cross over temperature" encountered in dissipative tunneling
(see eq.(3.7) of \cite{graolwei}), which will be commented on in the
next chapter. For illustrative purpose we may look once more at the
limiting case of zero friction for which one gets $T_0=\hbar
\varpi/(2\pi)$ and thus $2\,T_0=T_c$. The feature of the critical
temperature $T_c$ being larger than the $T_0$ is generally correct.

It is interesting to note that such a minimal temperature below which
motion can no longer be treated in a harmonic approximation is found
also in two extensions of the SPA, called P(erturbed)SPA in
\cite{attalha} and C(orrelated)SPA in \cite{roscan}. Both approaches
are very similar to one another, in the sense that small amplitude
vibrations are considered on top of or around the "static path"
approximation to the functional integral --- and are thus analogous in
spirit to our locally harmonic approximation. Being based on a path
integral formulation, it may be not very surprising that in this method
the minimal temperature turns out to be identical to $T_0$. It must be
said, however, that both in PSPA as well as in CSPA {\it no damping} of
the modes is considered. Within our time dependent formulation the
latter feature is made possible through considerations of coupling of
1p-1h-correlations to more complicated ones. In this sense the two body
interaction effectively goes beyond the separable form given in
(\ref{twobodham}), which just refers to the one generating collective
motion.  Certainly, more work will be needed to make the correspondence
of these different procedures more apparent.

\section{Summary and Conclusions}

In this paper we have reviewed a method which allows one to describe
the dynamics of damped motion in collective phase space. The main
emphasize was put on the evaluation of diffusion coefficients
$D_{\mu\nu}$ which govern motion of the second moments both for stable
as well as for unstable modes. The latter case was handled by a
suitable analytic continuation of the (quantal) fluctuation dissipation
theorem. To this end some elementary properties of linear response
functions had to be generalized, too.

From the numerical results it became evident that, when applied to the
diffusion coefficients, this continuation looses its physical meaning
below a critical temperature $T_c$ below which even the diagonal
element $D_{pp}$ would become negative. On the other hand, for the
fluctuation dissipation theorem, as well as for some basic relations of
response theory, such a continuation is possible even for temperatures
down to a $T_0<T_c$. This $T_0$ was seen to be identical to $T_0=\hbar
\vert\om_+\vert/(2\pi)$ where $\om_+=-(\om_+)^*$ measures the position
of the pole along the positive imaginary axis which belongs to the
unstable mode. This $T_0$ is thus identical to the so called "cross
over temperature" \cite{graolwei} one encounters in the phenomenon of
"dissipative tunneling" and for undamped motion in the PSPA and
the CSPA  of \cite{attalha} and \cite{roscan}, respectively. 

This association to tunneling is by far not accidental. As mentioned
above, the equations of motion associated to the harmonic case can be
exploited for describing global motion within a locally harmonic
approximation. In the end this means to describe collective dynamics by
way of a transport equations like the one of Kramers, with the only
difference of generalizing the classical Einstein relation to the
proper quantal form. This implies to have two diffusion coefficients
$D_{qp}$ and $D_{pp}$ instead of the single one $D_{pp}=\gamma \,T$.
Applying this scheme to evaluate the rate of decay out of a potential
minimum one finds \cite{hofthoming} the formula $R=f_{\rm Q} R_{\rm K}$
known from such model cases for which an application of functional
integrals becomes feasible. Compared to this latter situation, the
derivation sketched here has both an advantage as well as a
disadvantage. The advantage can be found in the fact that this locally
harmonic approximation can be applied to cases for which the coupling
between the collective degree of freedom and the rest is too big to
allow for the simplification of linearizing in the "bath variables",
with the latter simply being represented by oscillators with {\it
fixed} inertias and stiffnesses. On the other hand the description in
terms of transport equations of the type mentioned is valid only for
$T>T_c>T_0$. In this sense this critical temperature $T_c$ indicates a
breakdown of the locally harmonic approximation in general.

In this paper it has been emphasized that the construction of the basic
equations of motion is possible in self-consistent manor. By this
notion we mean that the collective degree of freedom is introduced to
represent the dynamical microscopic quantity one wants to study. This
is necessary for all systems which per se exist of identical particles,
like it is the case for nuclei or the electron gas, to mention just
two prominent examples.

\vspace{1cm}

\noindent
{\bf Acknowledgments:}

\vspace{0.2cm}
\noindent
The authors would like to thank R. Hilton, R. Rossignoli and P. Ring
for interesting comments and gratefully acknowledge financial support
by the Deutsche Forschungsgemeinschaft as well as by the National
Science Foundation under the Grant No.~PY-9403666. 

\vspace{1cm}

\begin{appendix}

\setcounter{equation}{0}
\renewcommand{\theequation}{\mbox{\Alph{section}.\arabic{equation}}}

\section{Extension of linear response theory to unstable modes}
\label{extlinres}
In the presence of an unstable mode of frequency $\om_+=i\Gamma$ the relations 
between response function have to be modified as compared to the common 
situation of a stable system. As we are going to show in this Appendix,
some relations between response function and its dissipative part can
be retained by making use of deformed paths of integration in the
complex frequency plane, as encountered in the text for the contour
$\cal C$ of (\ref{rftodrfst}).

Causality requires that a physical quantity can be influenced by
external fields only from their values in the past.  The retarded
function describing such a response must thus vanish for negative
times. This may be achieved by an appropriate choice of the contour
$\cal C_+$ for that integral which defines the Fourier transformation
from the frequency representation back to time: For negative times, it
must be possible to close this integral in the complex frequency plane
without enclosing any pole  of the retarded response function
$\chi^{\rm R}(\om)$. This allows one to write:
\bel{invftresf}
+ 2i\Theta(+ t)\chi^{\prime\prime}(t)=\int_{\cal C_+}
{d\om\over 2\pi}e^{-i\om t}\chi^{\rm R}(\om) \, .
\ee
The contour $\cal C_+$ could for instance be chosen parallel to the
real axis if it only lies {\it  above} the  $\om_+=i\Gamma$ (being the
pole of $\chi^{\rm R}$ having the largest imaginary part).  In the case
of the advanced response, in (\ref{invftresf}) all $+$ signs are to be
replaced by $-$ signs, with the contour $\cal C_-$ lying below all
poles of the advanced function.  In the following we will write both
versions in combined form, using an index X to distinguish between
R(etarded) and the A(dvanced).  Whenever the possible signs $\pm$ will
appear, the upper one will relate to the retarded, the lower one to the
advanced response.

To obtain the equivalent of the well known relation between the
frequency representations of response function and its dissipative part,
one may start with the inversion of (\ref{invftresf}),
\bel{ftresf}
\chi^{\rm X}(\om)=\pm\int dt e^{i\om t} 2i\Theta(\pm t)
\chi^{\prime\prime}(t)\qquad |\IM\om| > \Gamma \, ,
\ee
which is restricted to frequencies whose imaginary part is larger
(smaller) than  $\Gamma$. In a second step, one may combine
(\ref{ftresf}) for retarded and advanced response to find an expression
for the dissipative part of the response function,
\bel{dissrf}
\chi^{\prime\prime}(t)={1\over 2i}\left(
\int_{\cal C_+}{d\om\over 2\pi}e^{-i\om t}\chi^{\rm R}(\om)-
\int_{\cal C_-}{d\om\over 2\pi}e^{-i\om t}\chi^{\rm A}(\om)
\right)\, .
\ee
Under the assumption that retarded and advanced response do not share
any common pole both contours ${\cal C}_\pm$ may be deformed to a
common one $\cal C$ without changing the value of each integral.  One
only has to ensure that $\cal C$ has all poles of the retarded response
on one side and those of the advanced one on the other. For a stable
system, $\cal C$ can be taken along the real axis and the formulas
presented here turn to the ones known from the common literature.

The final step is to insert the dissipative part (\ref{dissrf}) into
(\ref{ftresf}) and to exchange the order of integration for time and
frequency. Defining the analytic continuation (with respect to $\om$)
of the dissipative part of the response function by
\bel{gendrf}
\chi^{\prime\prime}(\om)={1\over 2i}(\chi^{\rm R}(\om)-
                                        \chi^{\rm A}(\om))
\ee
one obtains the familiar form 
\bel{rftodrf}
\chi^{\rm R}(\om)=\int_{\cal C}{d\Om\over\pi}
{\chi^{\prime\prime}(\Om)\over\Om-\om}\qquad
{\rm for}\,\,  \IM\,\om > 0\,\, {\rm and}\,\, \om\ne i\Gamma
\ee
This equation is the key relation for obtaining the integral representation
(\ref{bcoreqinv})  for the fluctuation dissipation theorem from the
analytic continuation of its representation (\ref{eqqqserie}).

\section{The modified oscillator of the Drude regularization}
\label{sundrudereg}

In (\ref{drudegom}) a frequency dependent friction coefficient was
introduced to render some previously diverging integrals well behaved.
The form chosen there, $\gamma(\om)= \gamma ( 1-i\om/\omd)^{-1}$, is
commonly associated to the Drude regularization.  Replacing the
friction coefficient in the oscillator response (\ref{resosc}) by this
function the related dissipative part changes to
\bel{drustrength}
\chi_{\rm osc}^{\prime\prime {\rm dru}} (\om ) = 
{1 \over 2 } {\om \left(\gamma(\om) + \gamma^*(\om)\right)
     \over M(\om^2 - \varpi^2)^2 + i \om (\gamma(\om) - \gamma^*(\om))
     (\om^2 - \varpi^2) +\gamma(\om)\gamma^*(\om)\om^2}
\ee
It is apparent that with increasing $\om$ the form on right hand side of
(\ref{drustrength}) drops faster than the one with constant friction.
Therefore, the $\chi_{\rm osc}^{\prime\prime {\rm dru}} (\om )$
simulates better a correct behavior of the true response function at large
$\om$, whereas the difference to the mere oscillator response function
is small in the regions of maximal strength. 

The modified response function has one more pole as given by
the new secular equation
\bel{dispreldru}
C-i\om\Big({C\over\omd}+\gamma\Big)-M\om^2+i{M\over\omd}\om^3 =0
\ee
Actually, its solutions depend on two parameters only, 
as can be seen by introducing $ x=i {\om \over \omd} $
and rewriting (\ref{dispreldru}) to
\bel{dispreldrualt}
-x^3+x^2-\Big({C\over M\omd^2}+{\gamma\over M\omd}\Big)x+{C\over M\omd^2}=0
\ee
This equation allows for an analytic solution, but it is more
convenient to just solve it numerically. It is, however, instructive
to evaluate them in first order in $\varpi/\varpi_D$, or, less lengthy,
to solve directly (\ref{dispreldru}) perturbatively.  Defining $\eta=
\gamma/(2M\varpi)$ and $\zeta=\varpi/\varpi_D$, one finds for the first
two solutions
\begin{eqnarray}\label{carsolper}
   \om^{(1)} & \approx \varpi\left(+\sqrt{{\rm sign}C - \eta^2}-i\eta
   +  2\eta\zeta 
    \left[ {1 - 2\eta_C^2 \over 1 - \eta_C^2}
    \sqrt{{\rm sign}C - \eta^2} -i 2\eta\right]  \right)  \cr\\
   \om^{(2)} & \approx \varpi\left(-\sqrt{{\rm sign}C - \eta^2}-i\eta
   +   2\eta\zeta 
    \left[- {1 - 2\eta_C^2 \over 1 - \eta_C^2}
    \sqrt{{\rm sign}C - \eta^2} -i 2\eta\right]  \right)  \cr
\end{eqnarray}
It is easily recognized that for $\eta \zeta \to 0$ they turn into the
solutions $\om_\pm$ for the common oscillator, with
$\om_{\pm}=\pm\sqrt{C/M-\gamma^2/(2M^2)}-i\gamma/(2M)$. For the third 
solution one finds:
\bel{thirdsol}
\om^{(3)}  = -i \omd (1 - 2\eta\zeta)
\ee
A few remarks are in order here. First of all, it is interesting to
note that it is not the ratio $\zeta$ itself which matters but the
combination with $\eta$. It is intuitively clear that for larger
damping the Drude frequency must be chosen larger.  Secondly, we see
that for a large $\omd$, at least, the third pole lies on the negative
imaginary axis.

Let us finally just write down an expression for the full response
function which serves as starting point for expressing the fluctuations
in terms of the $\Psi-$function:
\bel{chioscdru}
\chidru= 
{-1\over M}{\om+i\omd\over (\om-\pol{1})(\om-\pol{2})(\om-\pol{3}) }
\ee
where $\om^{(k)}$ are the solutions of (\ref{dispreldru}).

\end{appendix}

\end{document}